\documentclass[12pt]{article}

\usepackage{array,dsfont} 
\usepackage{epsfig}
\usepackage{amssymb}
\usepackage{graphics,graphpap}

\renewcommand{\thefootnote}{\fnsymbol{footnote}}

\begin{document}

\begin{titlepage}
\begin{flushright}\begin{tabular}{l}
IPPP/06/75\\
DCPT/06/150
\end{tabular}
\end{flushright}
\vskip1.5cm
\begin{center}
   {\Large \bf\boldmath $|V_{ub}|$ from UTangles and 
    $B\to\pi \ell\nu$}
    \vskip2.5cm {\sc
Patricia Ball\footnote{Patricia.Ball@durham.ac.uk}
}
  \vskip0.5cm
{\em         IPPP, Department of Physics,
University of Durham, Durham DH1 3LE, UK }\\
\vskip2.5cm 


\vskip5cm

{\large\bf Abstract\\[10pt]} \parbox[t]{\textwidth}{
The angles of the CKM unitarity triangle are now known well enough to
allow a determination of its sides from global fits, with good
accuracy. Assuming that new physics does not affect the angles, 
UTfit and CKMfitter find $|V_{ub}|
= (3.50\pm 0.18)\times 10^{-3}$. Using this result, and new
high-precision data on the spectrum of $B\to\pi \ell\nu$ decays from BaBar, we
find $|V_{ub}| f_+(0) = (9.1\pm 0.7)\times 10^{-4}$ and $f_+(0) = 0.26\pm
0.02$, with $f_+$ the $B\to\pi$ weak transition form factor. 
These results are {\em completely model-independent}. 
We compare them to theoretical calculations from lattice and QCD 
sum rules on the light-cone. 
}

\end{center}
\end{titlepage}

\setcounter{footnote}{0}
\renewcommand{\thefootnote}{\arabic{footnote}}

\newpage

Ever since the first experimental observation of $b\to u$ transitions 
by the ARGUS collaboration 
in 1989 \cite{ARGUS}, the determination of $|V_{ub}|$ has been 
one of the major challenges for both experimental and theoretical B
physics. While initially exclusive transitions, in particular $B\to\pi
e \nu$, were considered as the most promising ones, the realisation
that inclusive decays can be calculated using heavy quark expansion
\cite{manifesto}, with (seemingly) controlled theoretical
uncertainties, has spurred a number of very impressive theoretical
works, culminating in the calculation of decay spectra based solely
on first principles (dressed gluon exponentiation [DGE]) 
\cite{einan} or using additional information from
other inclusive decays ($b\to s\gamma, b\to c \ell \nu$) 
in order to extract the
relevant non-perturbative quantities (BLNP) \cite{neubert}. At the same time,
the experimental measurement of inclusive $b\to u \ell\nu$ transitions
made major progress since the first measurements at ARGUS
and is now in a mature state. The results
for $|V_{ub}|$ determined in this way are collected and averaged by
the Heavy Flavour Average Group (HFAG) \cite{HFAG} and currently 
(November 06) read 
\begin{eqnarray}
|V_{ub}|_{\rm incl,DGE}^{\rm HFAG} &=& (4.46\pm 0.20({\rm exp})\pm
 0.20({\rm ext}))\times 10^{-3}\,,\nonumber\\
|V_{ub}|_{\rm incl,BLNP}^{\rm HFAG} &=& (4.49\pm 0.19({\rm exp})\pm
 0.27({\rm ext}))\times 10^{-3}\,,\label{VubHFAG}
\end{eqnarray}
where the first error is experimental (statistical and systematic) and
the second external (theoretical and parameter uncertainties). Both
results are in perfect agreement.

At the same time, $|V_{ub}|$ can also be determined in a more
indirect way, based on global fits of the unitarity triangle (UT),
using only input from various CP violating observables which are
sensitive to the angles of the UT. 
Following the UTfit collaboration, we call the corresponding fit of UT
parameters UTangles. 
To be precise, the information entering UTangles comes from the
following non-leptonic
decays: $B\to \pi\pi$, $B\to \pi\rho$ and
$B\to \rho\rho$ which yield the angle $\alpha$ 
\cite{alpha}; $B\to D^{(*)}\, K^{(*)}$ decays yielding
$\gamma$ \cite{gamma}; $2\beta+\gamma$ comes from
time-dependent asymmetries in $B \to D^{(*)} \pi (\rho)$
decays~\cite{2bpg} and $\cos 2 \beta $ from $B^0_d \to J/\psi
K^{*0}_S$~\cite{cos2b}; $\beta$ is determined 
from $B \to D^0\pi^0$~\cite{D0p0} and,
finally, $\sin 2 \beta$ from the ``golden mode'' $B^0_d \to J/\psi
K_S$~\cite{sin2b}. Both the UTfit \cite{UTfit} and the
CKMfitter collaboration \cite{CKMfitter,private} find
\begin{equation}\label{VubUT}
|V_{ub}|_{\rm UTangles}^{\rm UTfit,CKMfitter} = (3.50\pm 0.18)\times
 10^{-3}\,.
\end{equation}

The discrepancy between (\ref{VubHFAG}) and (\ref{VubUT}) starts to
become significant. One interpretation of this result is that there is
new physics (NP) in $B_d$ mixing which impacts the value of $\sin 2\beta$
from $b\to ccs$ transitions, the angle measurement with the
smallest uncertainty. The value of $|V_{ub}|$ in (\ref{VubHFAG}) implies
\begin{equation}
\left.\beta\right|_{|V_{ub}|_{\rm incl}^{\rm HFAG}} = (26.9\pm
2.0)^\circ\quad \longleftrightarrow \quad\sin 2\beta = 0.81\pm 0.04\,,
\end{equation}
using the recent Belle result $\gamma=(53\pm 20)^\circ$ from the
Dalitz-plot analysis of the tree-level process $B^+\to D^{(*)}K^{(*)+}$ 
\cite{Bellegamma}.\footnote{ We use the Belle measurement rather than
  that from BaBar, $\gamma=(92\pm 44)^\circ$ \cite{BaBargamma},
  because the uncertainty of the latter is too large to allow any
  meaningful statement. At present, HFAG does not provide an average
  of the BaBar and Belle measurements.} 
  This value disagrees by more than $2\sigma$ 
with the HFAG average for $\beta$ from 
$b\to ccs$ transitions, $\beta = (21.2\pm 1.0)^\circ$ ($\sin 2\beta =
0.675\pm 0.026$). The difference
  of these two results indicates the possible presence of a NP phase in
  $B_d$ mixing, $\phi_d^{\rm NP} \approx -10^\circ$.
This interpretation of the experimental situation 
is in line with that of Ref.~\cite{newphase}.
An alternative 
interpretation is that there is actually no or no significant NP in the
mixing phase of $B_d$ mixing, a scenario compatible with the MFV
hypothesis \cite{MFV}, but that the uncertainties in either
UTangles or inclusive $b\to u\ell\nu$ transitions (experimental and
theoretical) or both are
underestimated and that (\ref{VubHFAG}) and (\ref{VubUT}) actually do
agree.
In either case, the present situation calls for a critical re-assessment
of both UTangles and the inclusive analysis 
and for an independent 
determination of $|V_{ub}|$ from other sources. The aim of
this letter is to provide such a determination from exclusive $B\to\pi
\ell\nu$ decays, based on theoretical (lattice and QCD sum rule)
calculations and recent new data published by BaBar.

As for $B\to\pi\ell\nu$, the primary observable is the branching
ratio, for which HFAG quotes, combining charged and neutral $B$
decays using isospin symmetry \cite{HFAG}, 
\begin{equation}\label{4}
{\cal B}(\bar B^0\to \pi^+\ell^- \bar \nu_\ell) = (1.37\pm 0.06({\rm
  stat})\pm 0.06{\rm (syst)})\times 10^{-4}\,.
\end{equation}
The extraction of $|V_{ub}|$ from this measurement requires a
theoretical calculation of the hadronic matrix element
\begin{equation}
\langle \pi(p_\pi)| \bar u \gamma_\mu b | B(p_\pi+q)\rangle = \left(
2p_\pi{}_\mu  + q_\mu - q_\mu\,\frac{m_B^2-m_\pi^2}{q^2}\right) f_+(q^2) + 
 \frac{m_B^2-m_\pi^2}{q^2}\, q_\mu \,f_0(q^2)\,,
\end{equation}
where $q_\mu$ is the momentum of the lepton pair, 
with $m_{\ell}^2\leq q^2\leq
(m_B-m_\pi)^2=26.4\,$GeV$^2$. $f_+$ is the dominant form factor,
whereas $f_0$ enters only at order $m_{\ell}^2$ and can be neglected
for $\ell=e,\mu$. The spectrum in $q^2$ is then given by
\begin{equation}
\frac{d\Gamma}{dq^2}\,(\bar B^0\to \pi^+ \ell^- \bar\nu_\ell) = \frac{G_F^2
  |V_{ub}|^2}{ 192 \pi^3 m_B^3}\,\lambda^{3/2}(q^2) |f_+(q^2)|^2\,,
\end{equation}
where $\lambda(q^2) = (m_B^2+m_\pi^2-q^2)^2 - 4 m_B^2
m_\pi^2$ is the phase-space factor. The calculation of $f_+$ has
been the subject of numerous papers; the current state-of-the-art
methods are unquenched lattice simulations \cite{FNALno,HPQCD} and 
QCD sum rules on the light-cone (LCSRs) \cite{otherLCSR,BZ04}.
A particular challenge for any
theoretical calculation is the prediction of the {\em shape} of
$f_+(q^2)$ for all physical $q^2$: 
LCSRs effectively involve\footnote{This is
  not to say that LCSRs are a power expansion in $m_b/(2E_\pi)$, which
  is not a small parameter. Rather, the order parameter $1/m_b$ in the
  twist expansion of the LCSR for the form factor at $q^2=0$, see 
  Refs.~\cite{VtdVts}, becomes $1/(2E_\pi)$ for $q^2>0$,
  for contributions of twist 4 and higher.} 
the parameter $m_b/(2E_\pi)$ and become less
reliable for small $E_\pi$, i.e.\ large $q^2$. Lattice
calculations, on the other hand, are to date most reliable for small
$E_\pi$, although this is expected to change in the future
with the implementation of ``moving NRQCD'', i.e.\ a non-relativistic
description of the $b$ quark in a moving frame of reference (instead
of its rest frame) \cite{moving}. 
Hence, until very recently, the prediction of the
$B\to\pi \ell\nu$ decay rate necessarily involved an extrapolation of
the form factor, either to large or to small $q^2$. If, on the other
hand, the $q^2$ spectrum was known from experiment, the shape of $f_+$
could be constrained, allowing an extension of the LCSR and lattice
predictions beyond their region of validity. A first study of
the impact of the measurement of the $q^2$ spectrum in 5 bins in
$q^2$ by the BaBar collaboration \cite{bab} on the
shape of $f_+$ was presented in Ref.~\cite{BZvub}; in view of the
limited accuracy of the data available in 2005 the only firm
conclusion that could be drawn in \cite{BZvub} was that the simplest
possible parametrisation of the form factor by a simple pole at 
$q^2=m_{B^*}^2$, assuming
dominance of the $B^*(1^-)$ meson, is disfavoured.
The situation has
improved dramatically in summer 2006 with the publication of (preliminary)
high-precision data of the $q^2$ spectrum by the BaBar collaboration
\cite{BaBarSL}, with 12 bins in $q^2$ and full statistical and
systematic error correlation matrices.\footnote{The spectrum has been
  measured previously, by BaBar, CLEO and Belle
  \cite{bab,earlier}, in a smaller number of $q^2$ bins. As the new BaBar
  data are more precise, and the correlation of uncertainties is
  unknown for the earlier measurements, we do not include them in our 
analysis.} These data allow one to fit the
form factor to various parametrisations and determine the value of
$|V_{ub}| f_+(0)$. As it turns out, the results from all but the
simplest parametrisation agree up to tiny discrepancies which suggests
that the resulting value of $|V_{ub}| f_+(0)$ is {\em truly model-independent}.

There are four parametrisations of $f_+$ which are frequently used in
the literature. All but one of them include the essential feature that
$f_+$ has a pole at $q^2=m_{B^*}^2$; as $B^*(1^-)$ is a narrow
resonance with $m_{B^*}=5.325\,{\rm GeV}<m_B+m_\pi$, 
it is expected to have a
distinctive impact on the form factor. The parametrisations are:
\begin{itemize}
\item[(i)] Becirevic/Kaidalov (BK) \cite{BK}:
\begin{equation}\label{BK}
f_+(q^2) = \frac{f_+(0)}{\left(1-q^2/m_{B^*}^2\right)
  \left(1-\alpha_{\rm BK}\,q^2/m_{B}^2\right)}\,,
\end{equation}
where $\alpha_{\rm
  BK}$ determines the shape of $f_+$ and $f_+(0)$ the
  normalisation;
\item[(ii)] Ball/Zwicky (BZ) \cite{BZ04}:
\begin{equation}\label{BZ}
f_+(q^2) = f_+(0)\left(\frac{1}{1-q^2/m_{B^*}^2} + \frac{r
  q^2/m_{B^*}^2}{\left(1-q^2/m_{B^*}^2\right)\left(1-
\alpha_{\rm BZ}\,q^2/m_{B}^2\right)} \right),
\end{equation}
with the two shape parameters $\alpha_{\rm BZ}$, $r$ and the
normalisation $f_+(0)$; BK is a variant of BZ with $\alpha_{\rm BK} :=
\alpha_{\rm BZ} = r$;
\item[(iii)] the AFHNV parametrisation of Ref.~\cite{flynn}, based on
  an $(n+1)$-subtracted Omnes respresentation of $f_+$:
\begin{eqnarray}
f_+(q^2) \stackrel{n\gg 1}{=} \frac{1}{s_{th}-q^2}\,\prod_{i=0}^n \left[
  f_+(q_i)^2 (s_{th} - q_i^2)\right]^{\alpha_i(q^2)}\,,\\
\mbox{with}\quad \alpha_i(s) = \prod_{j=0,j\neq i}^n
  \frac{s-s_j}{s_i-s_j}\,, \quad s_{th} = (m_B+m_\pi)^2\,;
\end{eqnarray}
this parametrisation assumes that $f_+$ has {\em no} poles for $q^2<s_{th}$;
the shape parameters are $f_+(q_i^2)/f_+(q_0^2)$ with
$q_{0,\dots n}^2$ the subtraction points; following \cite{flynn}, 
we choose evenly spaced
$q_i^2 = q^2_{\rm  max} i/n$; again the normalisation is
given by $f_+(0)$; the assumption of no $B^*$ pole is likely to mostly
impact the form factor at large $q^2$;
\item[(iv)] the BGL parametrisation based on analyticity of $f_+$
  \cite{disper}: 
\begin{eqnarray}
f_+(q^2) & = & \frac{1}{P(t) \phi(q^2,q_0^2)}\,\sum_{k=0}^\infty
a_k(q_0^2) [z(q^2,q_0^2)]^k\,,\label{disper}\\
\mbox{with}\quad z(q^2,q_0^2) & = & \frac{\{(m_B+m_\pi)^2 - q^2\}^{1/2}
- \{(m_B+m_\pi)^2 - q_0^2\}^{1/2}}{ \{(m_B+m_\pi)^2 - q^2\}^{1/2}
+ \{(m_B+m_\pi)^2 - q_0^2\}^{1/2}}
\end{eqnarray}
with $\phi(q^2,q_0^2)$ as given in \cite{disper}. The ``Blaschke'' factor
$P(q^2) = z(q^2,m_{B^*}^2)$ accounts for the $B^*$ pole. The expansion
parameters $a_k$ are constrained by unitarity to fulfill $\sum_k a_k^2
\leq 1$. $q_0^2$ is a
free parameter that can be chosen to attain the tightest possible
bounds, and it defines $z(q_0^2,q_0^2) = 0$; $|z|<1$ for
$q_0^2<(m_B+m_\pi)^2$. 
The series in (\ref{disper}) provides a systematic expansion in the
small parameter $z$, which for practical purposes has to be truncated
at order $k_{\rm max}$. We let data decide where to truncate and do
a $\chi^2_{\rm min}$ analysis for increasing $k_{\rm max}$ till an
absolute minimum of $\chi^2_{\rm min}$ is reached. 
The shape parameters are then given by
$\{a_k\} \equiv \{ \tilde a_k\}\times\mbox{const.}$ 
and we choose $\mbox{(const.)}^2 =
  \sum_0^{k_{\rm max}} a_k^2$, which implies $\sum_0^{k_{\rm max}}
 \tilde{a}_k^2=1$, so
that the $\tilde a_k$ can be parametrised by
  generalised $k_{\rm max}+1$ dimensional spherical polar angles. For
  $k_{\max}=2$ we choose
\begin{equation}\label{as}
\tilde a_0 = \cos \theta_1,\quad \tilde a_1 = \sin\theta_1\cos
\theta_2,\quad \tilde a_2 = \sin\theta_1\sin\theta_2,
\end{equation}
and for $k_{\rm max}=3$
\begin{equation}
\tilde a_0 = \cos \theta_1,\quad \tilde a_1 = \sin\theta_1\cos
\theta_2,\quad \tilde a_2 = \sin\theta_1\sin\theta_2\cos\theta_3,\quad
\tilde a_3 = \sin\theta_1\sin\theta_2\sin\theta_3.
\end{equation}
We then minimize $\chi^2$ in $\theta_{i}$ for the shape of $f_+$,
for two choices of $q_0^2$: 
\begin{itemize}
\item[(a)] $q_0^2 = (m_B+m_\pi) (\sqrt{m_B}-\sqrt{m_\pi})^2 =
20.062\,{\rm GeV}^2$, which minimizes the possible values of $z$,
$|z|<0.28$, and hence also minimizes the truncation error of the series in
(\ref{disper}) across all $q^2$; the minimum $\chi^2$ is reached for 
$k_{\rm max} = 2$;
\item[(b)] $q_0^2 = 0\,{\rm GeV}^2$ with
$z(0,0)=0$ and $z(q^2_{\rm max},0)= -0.52$, which minimizes the
truncation error for small and moderate $q^2$ where the data are most
constraining; the minimum $\chi^2$ is reached for $k_{\rm max} = 3$. 
\end{itemize}
\end{itemize}
The advantage of BK and BZ is that they are both intuitive and simple;
they are obtained from the dispersion relation for $f_+$,
\begin{equation}\label{eq:disper}
f_+(q^2) = \frac{{\rm Res}_{q^2=m_{B^*}^2} f_+(q^2)}{q^2-m_{B^*}^2} +
\frac{1}{\pi} \,\int_{(m_B+m_\pi)^2}^\infty dt\,\frac{{\rm
    Im}\,f_+(t)}{t-q^2-i \epsilon}\,,
\end{equation}
by replacing the second term on the right-hand side by an effective pole.
However, they cannot easily be extended to include more
parameters. AFHNV, on the other hand,
is based on a completely different approach, so it is interesting to
compare the best-fit results with those from the other approaches; its
shortcoming is the failure to include the $B^*$ pole, which is
possible, but difficult, see Ref.~\cite{flynn}. Finally, BGL offers a
systematic expansion whose accuracy can be adapted to that of the data
to be fitted, so we choose it as our default parametrisation.

We determine the best-fit parameters for all four parametrisations
from a minimum-$\chi^2$ analysis. 
Our results are given in Tabs.~\ref{tab1} and \ref{tab2}.
\begin{table}[tbp]
\renewcommand{\arraystretch}{1.3}
\addtolength{\arraycolsep}{3pt}
$$
\begin{array}{l||l|l}
& |V_{ub}| f_+(0) & \mbox{Remarks}\\\hline
{\rm BK} & (9.3\pm 0.3\pm 0.3)\times 10^{-4}
         & \chi^2_{\rm min}=8.74/11\,{\rm dof}\\
         & & \alpha_{\rm BK} = 0.53\pm 0.06\\\hline
{\rm BZ} & (9.1\pm 0.5\pm 0.3)\times 10^{-4}
         & \chi^2_{\rm min}=8.66/10\,{\rm dof}\\
         & & \alpha_{\rm BZ} 
= 0.40^{+0.15}_{-0.22},\,r = 0.64^{+0.14}_{-0.13} \\\hline
{\rm BGLa} & (9.1\pm 0.6 \pm 0.3)\times 10^{-4}
          & \chi^2_{\rm min}=8.64/10\,{\rm dof}\\
         && q_0^2 = 20.062\,{\rm GeV}^2\\
         && \theta_1 = 1.12^{+0.03}_{-0.04},\,\theta_2 = 4.45\pm 0.06\\\hline
{\rm BGLb} & (9.1\pm 0.6 \pm 0.3)\times 10^{-4}
          & \chi^2_{\rm min}=8.64/9\,{\rm dof}\\
         && q_0^2 = 0\,{\rm GeV}^2\\
         && \theta_1 = 1.41^{+0.02}_{-0.03},\,\theta_2 = 3.97\pm
  0.10\,,
         \theta_3 = 5.11^{+0.67}_{-0.39}\\\hline
{\rm AFHNV} & (9.1\pm0.3\pm0.3)\times 10^{-4} & \chi^2_{\rm min} = 8.64/8\,{\rm
  dof}\\
&& f_+(q^2_{\rm max}\cdot \{1/4,2/4,3/4,4/4\})/f_+(0)\\
&&  = \{1.54\pm
0.07,1.54\pm 0.11,5.4\pm 0.4,26\pm 11\}\\\hline\hline
{\rm SCET} & (8.0\pm 0.4)\times 10^{-4} & \mbox{using the method of
  Ref.~\cite{grin}}
\end{array}
$$
\caption[]{Model-independent results 
  for $|V_{ub}| f_+(0)$ using the BaBar data for the
  spectrum \cite{BaBarSL} and the HFAG average for the total branching
  ratio ${\cal B}(B\to\pi \ell\nu)=(1.37\pm 0.08)\times 10^{-4}$
  \cite{HFAG}. The results are obtained using different parametrisations
  of the form factor $f_+(q^2)$: 
Becirevic/Kaidalov (BK) \cite{BK}, Ball/Zwicky (BZ)
  \cite{BZ04}, Boyd/Grinstein/Lebed (BGL) \cite{disper} and the Omnes
  representation of Ref.~\cite{flynn} (AFHNV). The first error is induced by
  the uncertainties of the parameters determining the shape of $f_+$;
  these parameters are given in the right column (our result for
  $\alpha_{\rm BK}$ coincides with that obtained in \cite{BaBarSL}). The
  second error comes from the uncertainty of the branching ratio.
  We also give the corresponding result obtained
  from $B\to\pi\pi$ decays using SCET \cite{grin} (with
  $\gamma=(53\pm 20)^\circ$); the error is purely
  experimental. 
}\label{tab1}
\end{table}
In Tab.~\ref{tab1} we give the results for $|V_{ub}|f_+(0)$ obtained from
fitting the various parametrisations to the BaBar data for the
normalised partial branching fractions in 12 bins of $q^2$:
$q^2\in \{[0,2],[2,4],[4,6],[6,8],[8,10],[10,12],$ $[12,14],[14,16],[16,18],
[18,20],[20,22],[22,26.4]\}\,{\rm GeV}^2$; the absolute 
normalisation is gi\-ven by the HFAG average of the
semileptonic branching ratio, Eq.~(\ref{4}). 
It is evident that good values of $\chi^2_{\rm min}$ are
obtained for all parametrisations. We have also determined $\chi^2$
for the (parameter-free) simple-pole/vector-dominance
parametrisation $f_+\propto 1/(1-q^2/m_{B^*}^2)$ and find $\chi^2 =
45.3$ implying that this shape is largely incompatible with data,
which confirms the result of Ref.~\cite{BZvub}. The central values of
$|V_{ub}|f_+(0)$ agree for all parametrisations with more than one
shape parameter, i.e.\ all parametrisations except the simplest one,
BK. The uncertainty induced by the shape parameters is largest for the
BGL parametrisation. As our final result we quote
\begin{equation}\label{16}
|V_{ub}|f_+(0) = (9.1\pm 0.6({\rm shape})\pm 0.3({\rm branching~
 ratio}))\times 10^{-4}
\end{equation}
and choose BGLa as default parametrisation with best $\chi^2_{\rm
  min}$ for a minimum number of parameters.
We would like to stress that this result is {\em completely
  model-independent}, and also independent of the value of $|V_{ub}|$; it 
relies solely on the experimental data for $B\to\pi\ell\nu$ from BaBar
 for the spectrum \cite{BaBarSL} and the HFAG average of the branching
  ratio, Eq.~(\ref{4}). Using the two competing results for
  $|V_{ub}|$, (\ref{VubHFAG}) and (\ref{VubUT}), (\ref{16}) implies
\begin{equation}
\left.f_+(0)\right|_{\rm incl}=0.20\pm0.02\,,\qquad
 \left.f_+(0)\right|_{\rm UTangles}=0.26\pm0.02\,.
\end{equation}
We also give the result for $|V_{ub}|
f_+(0)$ obtained from non-leptonic $B\to\pi\pi$ decays using SCET
\cite{grin}. 
The method used in \cite{grin} to obtain a constraint on
  $|V_{ub}| f_+(0)$ from the decay rates for $B^\pm\to \pi^\pm \pi^0$
  and $B^0\to\pi^+\pi^-$ and the CP asymmetries of the latter decay
  is only valid at tree-level and to leading order in $1/m_b$;
  corrections to this relation are of order $\alpha_s$ and $1/m_b$ and
  apparently, according to the BaBar data, are of the order of
  15\%. It remains to be seen whether a calculation of these
  corrections in SCET is feasible. 

\begin{figure}[tb]
$$\epsfxsize=0.48\textwidth\epsffile{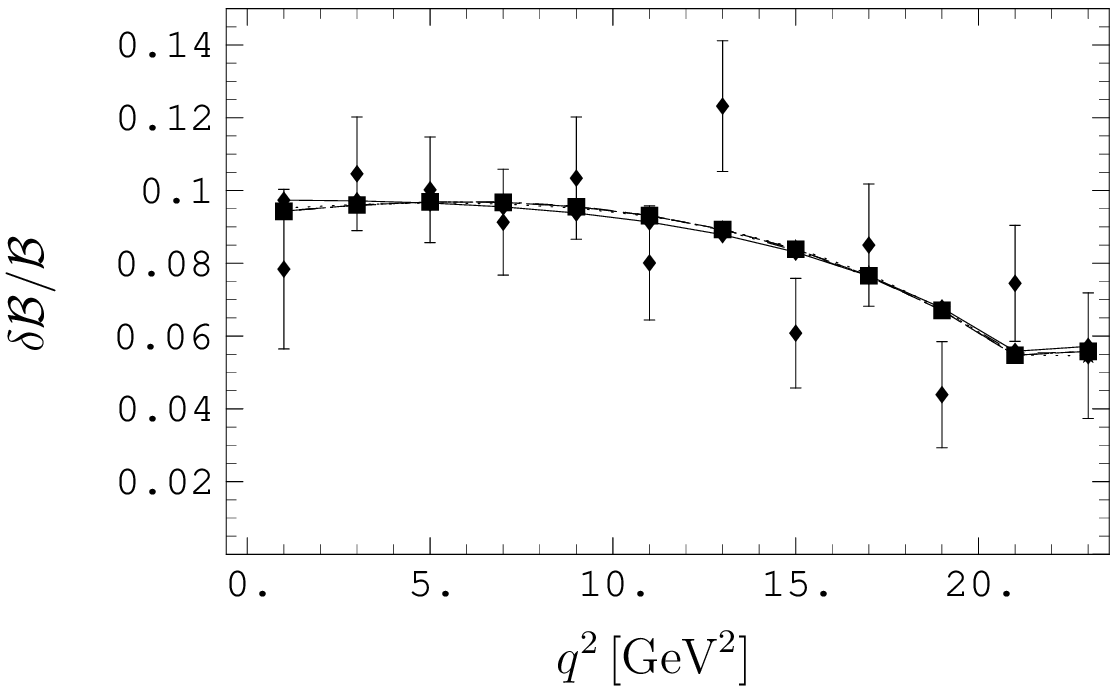}$$
\vspace*{-30pt}
\caption[]{Experimental data for the normalised branching ratio
  $\delta{\cal B}/{\cal B}$ per
  $q^2$ bin, $\sum \delta{\cal B}/{\cal B}=1$, and best
  fits. We have added statistical and systematic
  errors in quadrature.  The  lines are the best fit results for
  the five different parametrisations listed in Tab.~\ref{tab1}.
The increase in the last bin is due to
  the fact that it is wider than the others ($4.4\,{\rm GeV}^2$ vs.\
  $2\,{\rm GeV}^2$).}\label{fig1}
$$\epsfysize=5cm\epsffile{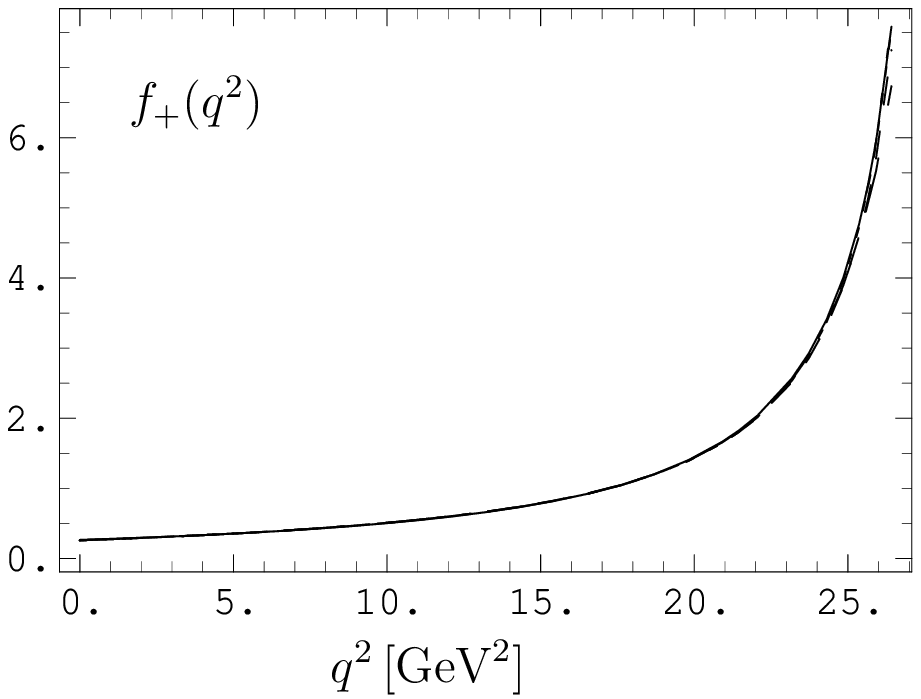}
\qquad\epsfysize=5cm\epsffile{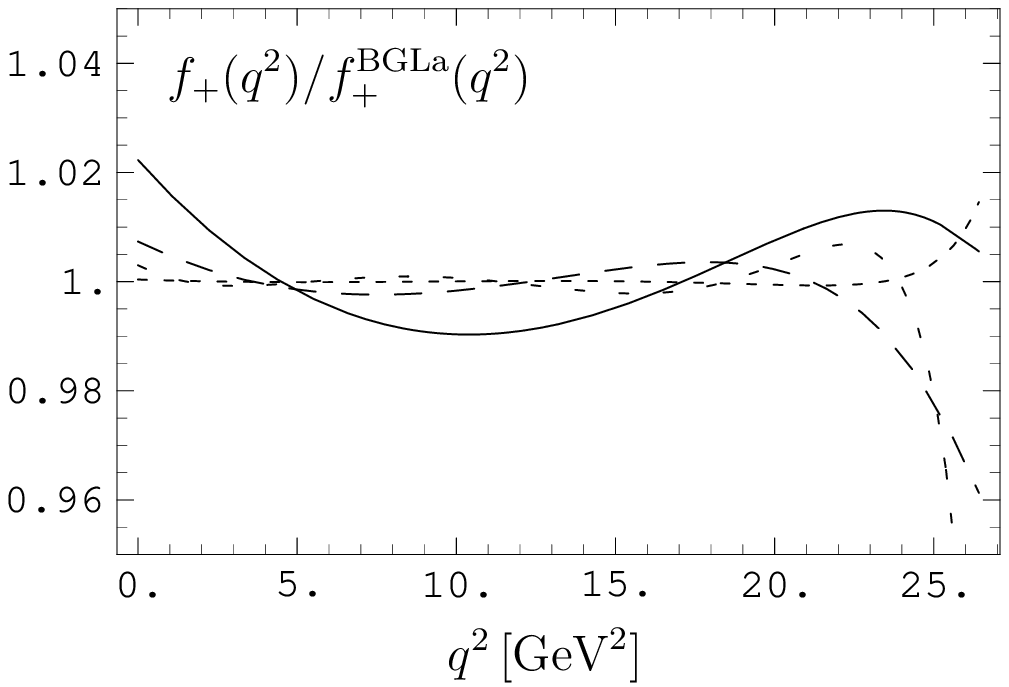} $$
\vspace*{-30pt}
\caption[]{Left panel: best-fit form factors $f_+$ as a function of
  $q^2$. The line is an overlay of all five parametrisations. 
Right panel: best-fit form factors normalised to BGLa. Solid
  line: BK, long dashes: BZ, short dashes: BGLb, short dashes with
  long spaces: AFHNV.}\label{fig2}
\end{figure}

In  Fig.~\ref{fig1} we show the best fit curves for all parametrisations
  together with the experimental data and error bars.
All fit curves basically coincide except for the BK parametrisation
which has a slightly worse $\chi^2_{\rm min}$. 
This can be easily understood because BK has only
one shape parameter which is not sufficient to describe the whole
spectrum. Still, BK gives, within errors, the same result for
$|V_{ub}| f_+(0)$ as the other parametrisations.
The situation becomes more
complicated, however, 
if one wants to fit lattice data obtained at large $q^2$ to
BK, as done, for instance, in Ref.~\cite{FNALno}; 
we will come back to that point below when
discussing lattice data. In Fig.~\ref{fig2} we show
the best-fit form factors themselves. The  curve in the left
panel is an overlay
of all five parametrisations; noticeable differences only occur for
large $q^2$, which is due to the fact that these points are 
phase-space suppressed in the spectrum and hence cannot be fitted with
high accuracy. In the right panel we graphically 
enhance the differences
between the best fits by normalising all parametrisations
to our preferred choice BGLa; for $q^2<25\,{\rm GeV}^2$, all best-fit form
factors agree within 2\%. It is also evident that BZ and AFHNV
yield too slow an increase for $q^2>25\,{\rm GeV}^2$, that is in close
proximity of the $B^*$ pole at $28.4\,{\rm GeV}^2$. For AFHNV this is
expected, as it features the pole at a slightly larger $q^2$,
$(m_B+m_\pi)^2 = 29.4\,{\rm GeV}^2$. In Tab.~\ref{tab2} we give explicit 
values for the best-fit form factors for various $q^2$.

\begin{table}[tbp]
\renewcommand{\arraystretch}{1.3}
\addtolength{\arraycolsep}{3pt}
$$
\begin{array}{l||l|l|l|l|l}
 & {\rm BK} & {\rm BZ} & {\rm BGLa} & {\rm BGLb} & {\rm AFHNV}\\\hline
f_+(0)     & 0.26\pm 0.01 & 0.26\pm 0.01 & 0.26\pm 0.02 & 0.26\pm 0.02
& 0.26\pm 0.01\\
f_+(5)     & 0.35\pm 0.01 & 0.35\pm 0.01 & 0.36\pm 0.01 & 0.36\pm 0.01 
& 0.36\pm 0.02\\
f_+(10)    & 0.50\pm 0.01 & 0.51\pm 0.01 & 0.51\pm 0.01 & 0.51\pm 0.01 
& 0.51\pm 0.02\\
f_+(15.23) & 0.80\pm 0.01 & 0.81\pm 0.02 & 0.81\pm 0.02 & 0.81\pm 0.02 
& 0.80\pm 0.03\\
f_+(15.87) & 0.86\pm 0.01 & 0.86\pm 0.02 & 0.86\pm 0.02 & 0.86\pm 0.02 
& 0.86\pm 0.04\\
f_+(16.28) & 0.90\pm 0.02 & 0.90\pm 0.02 & 0.90\pm 0.03 & 0.90\pm 0.03 
& 0.90\pm 0.04\\
f_+(17.34) & 1.01\pm 0.02 & 1.02\pm 0.03 & 1.01\pm 0.04 & 1.01\pm 0.04 
& 1.01\pm 0.04\\
f_+(18.39) & 1.15\pm 0.03 & 1.15\pm 0.05 & 1.15\pm 0.05 & 1.15\pm 0.06 
& 1.15\pm 0.06\\
f_+(18.58) & 1.18\pm 0.04 & 1.18\pm 0.05 & 1.18\pm 0.05 & 1.18\pm 0.06 
& 1.18\pm 0.06\\
f_+(19.45) & 1.33\pm 0.05 & 1.33\pm 0.07 & 1.32\pm 0.06 & 1.32\pm 0.08 
& 1.33\pm 0.08\\
f_+(20.51) & 1.56\pm 0.06 & 1.55\pm 0.09 & 1.55\pm 0.08 & 1.54\pm 0.11 
& 1.55\pm 0.10\\
f_+(21.56) & 1.86\pm 0.09 & 1.84\pm 0.12 & 1.84\pm 0.10 & 1.84\pm 0.15 
& 1.85\pm 0.15\\
f_+(24.09) & 3.21\pm 0.21 & 3.13\pm 0.29 & 3.17\pm 0.23 & 3.17\pm 0.39 
& 3.16\pm 0.53
\end{array}
$$
\vspace*{-10pt}
\caption[]{Results for $f_+(q^2)$ using the best fits collected in
  Tab.~\ref{tab1} and the UTangles value for $|V_{ub}|$,
  $(3.5\pm 0.18)\times 10^{-3}$. The errors refer to the
  fit of the various parametrisations to the data; the additional
  error induced by $|V_{ub}|$ is $\pm 5\%$ and that from the total
  branching ratio $\pm 3\%$.}\label{tab2}

\medskip

\renewcommand{\arraystretch}{1.3}
$$
\begin{array}{c|c||c|c||c|c}
q^2\,[{\rm GeV}]^2 
& \mbox{LCSRs \cite{BZ04}} & q^2\,[{\rm GeV}]^2 & \mbox{HPQCD \cite{HPQCD}} & 
q^2\,[{\rm GeV}]^2 & \mbox{FNAL \cite{grin}}\\\hline
0 & 0.26\pm 0.03 
& 15.23 & 0.649\pm 0.063 
& 15.87 & 0.799\pm 0.058
\\
2 & 0.29\pm 0.03
& 16.28 & 0.727\pm 0.064
& 18.58 & 1.128\pm 0.086
\\
4 & 0.33\pm 0.04
& 17.34 & 0.815\pm0.065
& 24.09 & 3.262\pm0.324
\\
6 & 0.38\pm0.05
& 18.39 & 0.944\pm 0.066
&
\\
8 & 0.44\pm 0.05
& 19.45 & 1.098\pm 0.067
&
\\
10 & 0.52\pm 0.06
& 20.51 & 1.248\pm0.097
&
\\
& 
& 21.56 & 1.554\pm0.156
&
\end{array}
$$
\caption[]{Theoretical predictions of $f_+(q^2)$ from LCSRs
  \cite{BZ04} and lattice \cite{HPQCD,grin}. 
  The errors quoted for HPQCD are combined
  statistical and chiral extrapolation errors.
  The FNAL numbers are quoted
  from Ref.~\cite{grin}, as we were unable to track down any
  publication of the FNAL group giving these numbers; the  error
  is statistical.}\label{tab3}
\end{table}

\begin{table}[tb]
\renewcommand{\arraystretch}{1.3}
\addtolength{\arraycolsep}{3pt}
$$
\begin{array}{l||l|l}
& \mbox{BK} & \mbox{BGLa}\\\hline
\mbox{LCSR}
& f_+(0)=0.26\pm 0.03\,,\quad \alpha_{\rm BK} = 0.63^{+0.18}_{-0.21}
& f_+(0)=0.26\pm 0.03\\
\mbox{Ref.~\cite{BZ04}}
& |V_{ub}| = (3.5\pm 0.6\pm 0.1)\times 10^{-4} & |V_{ub}| = (3.5\pm
0.4\pm 0.1)\times 10^{-4}\\
& |V_{ub}| f_+(0) = (9.0^{+0.7}_{-0.6}\pm 0.4)\times 10^{-4} &\\
\mbox{exp. input} & {\cal B}(B\to\pi \ell\nu)_{q^2\leq 16\,{\rm
    GeV}^2}
& {\cal B}(B\to\pi \ell\nu) \mbox{~and~BGLa}\\
&  = (0.95\pm 0.07)\times 10^{-4}
& \mbox{parameters from Tab.~\ref{tab1}}\\ \hline
\mbox{HPQCD} 
& f_+(0)=0.21\pm 0.03\,,\quad \alpha_{\rm BK} = 0.56^{+0.08}_{-0.11}
& f_+(0)=0.21\pm 0.03\\
\mbox{Ref.~\cite{HPQCD}} 
& |V_{ub}| = (4.3\pm 0.7\pm 0.3)\times 10^{-4} & |V_{ub}| = (4.3\pm
0.5\pm 0.1)\times 10^{-4}\\
& |V_{ub}| f_+(0) = (8.9^{+1.2}_{-0.9}\pm 0.4)\times 10^{-4} &\\
\mbox{exp. input} & {\cal B}(B\to\pi \ell\nu)_{q^2\geq 16\,{\rm
    GeV}^2}
& {\cal B}(B\to\pi \ell\nu) \mbox{~and~BGLa}\\
&  = (0.35\pm 0.04)\times 10^{-4} 
& \mbox{parameters from Tab.~\ref{tab1}}\\ \hline
\mbox{FNAL} 
& f_+(0)=0.23\pm 0.03\,,\quad \alpha_{\rm BK} = 0.63^{+0.07}_{-0.10}
& f_+(0)=0.25\pm 0.03\\
\mbox{Ref.~\cite{grin}} 
& |V_{ub}| = (3.6\pm 0.6\pm 0.2)\times 10^{-4} & 
|V_{ub}| = (3.7\pm 0.4\pm 0.1)\times 10^{-4}\\
& |V_{ub}| f_+(0) = (8.2^{+1.0}_{-0.8}\pm 0.3)\times 10^{-4} &\\
\mbox{exp. input} & {\cal B}(B\to\pi \ell\nu)_{q^2\geq 16\,{\rm
    GeV}^2}
& {\cal B}(B\to\pi \ell\nu)\mbox{~and~BGLa}\\
&   = (0.35\pm 0.04)\times 10^{-4} 
& \mbox{parameters from Tab.~\ref{tab1}}
\end{array}
$$
\caption[]{$|V_{ub}|$ and $|V_{ub}| f_+(0)$ from various theoretical
  methods. The column labelled BK gives the results obtained from a
  fit of the form factor to the BK parametrisation, and the column
  labelled BGLa that from a fit of $f_+(0)$ 
to the best-fit BGLa parametrisation from Tab.~\ref{tab1}. The first
  uncertainty comes from the shape parameters, the second from the
  experimental branching ratios;
  the latter are taken from HFAG \cite{HFAG}.
}\label{tab4}
\end{table}

As mentioned above, theoretical predictions for $f_+$ are available
from lattice calculations and LSCR and are
collected in Tab.~\ref{tab3}. The LCSR calculations \cite{BZ04} 
include twist 2 and 3 contributions to $O(\alpha_s)$ accuracy and twist-4
contributions at tree-level. The lattice calculations are unquenched 
with $N_f=2+1$ dynamical flavours, i.e.\
mass-degenerate $u$ and $d$ quarks and a heavier $s$ quark. These
quarks are described by an improved staggered quark action, which
allows a simulation much closer to the (physical) chiral limit than
with alternative actions. The two calculations differ in the treatment
of the $b$ quark: whereas HPQCD simulates it in
nonrelativistic QCD, FNAL employs a tadpole-improved clover action
with the Fermilab interpretation.  The obvious questions are (a) whether these
predictions are compatible with the experimentally determined shape of
the form factor and (b) what the resulting value of $|V_{ub}|$
is. In order to answer these questions, we fit the lattice and LCSR
form factors to the BK parametrisation and extract $|V_{ub}|$, for
lattice, from ${\cal B}(B\to\pi\ell\nu)_ {q^2\geq 16\,{\rm
  GeV}^2}$, and for LCSR from ${\cal B}(B\to\pi\ell\nu)_ {q^2\leq 16\,{\rm
  GeV}^2}$; the cuts in $q^2$ are imposed in order to minimise any
uncertainty from extrapolating in $q^2$. 
In our fits we treat the theory errors given in
Tab.~\ref{tab3} as uncorrelated and add another 12\% fully correlated
systematic error, both for LCSR and lattice predictions, which  
is the procedure followed by experimental and
lattice papers (with the exception of Ref.~\cite{HPQCD} where BZ is
used). 
The results are shown in the BK column of
Tab.~\ref{tab4}. Equipped with the experimental information on the form factor
shape, i.e.\ the BGLa parametrisation of Tab.~\ref{tab1},
we suggest a different procedure and 
perform a fit of the theoretical predictions to
this shape, with the normalisation as fit parameter. The corresponding
results are shown in the right column. Comparing these results, we
observe the following:
\begin{itemize}
\item $|V_{ub}|$ from LCSR and FNAL is in better agreement with the
  UTangles value (\ref{VubUT}) than that from inclusive decays
  (\ref{VubHFAG});  $|V_{ub}|$ from HPQCD agrees with (\ref{VubHFAG});
  the discrepancy between LCSR and (\ref{VubHFAG}) is at the $2\sigma$
  level, for FNAL it is slightly smaller;
\item the difference in results for the two parametrisations is
  strongest for FNAL, which is due to the small number of theory input
  points (3); the ``quality'' of the BK
  parametrisation can be measured by the result for $|V_{ub}| f_+(0)$
  which for LCSR and HPQCD perfectly agrees with the experimental 
  value (\ref{16}), whereas the central value for FNAL is a bit low;  
\item comparing the errors for $|V_{ub}|$ in both columns,
  it is evident that the main impact of the experimentally fixed
  shape, i.e.\ using the BGLa parametrisation of $f_+$,
  is a reduction of both theory and experimental errors; this is due to the
  fact that, once the shape is fixed, $|V_{ub}|$ can be determined
  from the full branching ratio with only 3\% experimental
  uncertainty, whereas the partical branching fractions in the BK
  column induce 4\% and 6\% uncertainty, respectively, for $|V_{ub}|$;
  the theory error becomes smaller because the errors on $f_+$ in
  Tab.~\ref{tab3} are still rather large, which implies errors
  on the shape parameter $\alpha_{BK}$ which are larger than those of
  the experimentally fixed shape parameters.
\end{itemize}
The main conclusion from this discussion is that both LCSR and FNAL
predictions for $f_+$ support the UTangles value for $|V_{ub}|$,
and differ at the $2\sigma$ level from the inclusive $|V_{ub}|$,
whereas HPQCD supports the inclusive result. Using the experimentally
fixed shape of $f_+$ in the analysis instead of fitting it to the
theoretical input points reduces both the theoretical and experimental
uncertainty of the extracted $|V_{ub}|$.

To summarize, we have presented a truly model-independent 
determination of the quantity $|V_{ub}| f_+(0)$ from the experimental data
for the spectrum of $B\to\pi\ell\nu$ in the invariant lepton mass
provided by the BaBar collaboration \cite{BaBarSL}; 
our result is given in (\ref{16}). We
have found that the BZ, BGL and AFHNV parametrisations of the form
factor yield, to within 2\% accuracy, the same results for
$q^2<25\,{\rm GeV}^2$. We then have used the best-fit BGLa shape of $f_+$ to
determine $|V_{ub}|$ using three different theoretical predictions for
$f_+$, QCD sum rules on the light-cone \cite{BZ04}, and the lattice
results of the HPQCD \cite{HPQCD} and FNAL collaborations
\cite{FNALno,grin}. The advantage of this procedure compared to that
employed in previous works, where the shape was determined from the
theoretical calculation itself, is a
reduction of both experimental and theoretical uncertainties of the
resulting value of $|V_{ub}|$. We have found that the LCSR and FNAL
form factors yield values for $|V_{ub}|$ which
agree with the UTangles result, but
differ, at the $2\sigma$ level, from the HFAG value
obtained from inclusive decays. The HPQCD form factor, on
the other hand, is compatible with both UTangles and the inclusive
$|V_{ub}|$. Our results show a certain preference for the UTangles
result for $|V_{ub}|$, disfavouring a new-physics scenario in $B_d$
mixing, and highlight the quite urgent need for a
re-analysis the inclusive case.

\subsection*{Acknowledgements}
This work was supported in part by the EU-RTN network  {\sc
  Flavianet}, contract No.\
MRTN-CT-2006-035482.

\end{document}